\documentclass[letterpaper,preprintnumbers,superscriptaddress,aps,nofootinbib,perprint,11pt]{revtex4}
\usepackage{CJK}\usepackage{amssymb}\usepackage[centertags]{amsmath}\usepackage{txfonts}
\usepackage{epsfig}\usepackage{bm}\usepackage{color}\usepackage{graphicx,graphics}
\usepackage{multirow}\usepackage{float}\usepackage{ulem}\usepackage{hyperref}
\usepackage{setspace}\usepackage{slashed}
\usepackage{booktabs}\usepackage{diagbox}\usepackage{epstopdf}
\hypersetup{colorlinks=true, citecolor=blue, linkcolor=blue,filecolor=black,urlcolor=blue}
\allowdisplaybreaks[4]

\begin{document}

\title{The hadron pair forward-backward asymmetry in the electron positron  annihilation process}

\author{Weihua Yang}
\affiliation{College of Nuclear Equipment and Nuclear Engineering, Yantai University,  Yantai, 264005, China }

\author{Chao Li}
\affiliation{School of Mathematics and Physics, Nanyang Institute of Technology,  Nanyang, 473000, China}

\begin{abstract}
The forward-backward asymmetry is an important measurable quantity which can enable independent determination of the neutral-current couplings of fermions. In this paper, we extend the definition of the asymmetry from the partonic level to the hadronic level by calculating this asymmetry of the hadron pair in the semi-inclusive electron position annihilation process. Semi-inclusive implies that a back-to-back jet is also measured in addition to the hadron pair. Due to the limitation of the factorization theorem, we calculate this process up to leading order twist-4 level by applying the collinear expansion formalism. After obtaining the differential cross section, we calculate the forward-backward asymmetries and show them in terms of the corresponding di-hadron fragmentation functions. Di-hadron fragmentation functions are introduced to describe the hadron pair productions in the fragmentation process. With available parameterization of the functions, we present a numerical estimation of the forward-backward asymmetry. On the flip side, measurements of the forward-backward asymmetry would give strict restrictions of parameterizations of the di-hadron fragmentation functions. We also present a numerical estimation of the twist-4 di-hadron fragmentation functions in order to illustrate their contributions.

\end{abstract}

\maketitle

\section{Introduction}

The spontaneously broken electroweak (EW) theory and the color gauge theory or quantum chromodynamics (QCD) are two basic components of the Standard Model (SM) which has achieved great success in describing the elementary particles and their interactions. Since the left- and right-handed fermions in the EW theory live in different representations of the fundamental gauge group, they have different couplings for the gauge bosons. As for $Z^0$, the difference leads to an asymmetry in the angular distribution of positively and negatively charged fermions (leptons and quarks) produced in $Z^0$ decays. This asymmetry, known as the forward-backward asymmetry~\cite{Wu:1984ik,Marshall:1988bk,Langacker:1995eqb,ALEPH:2005ab}, depends on the weak mixing angle and can enable independent determinations of the neutral-current couplings of these fermions. Asymmetries for known fermions have definite distributions with respect to center-of-mass energy. They can be measured with very precision at the $Z^0$ pole. Therefore, it is more interesting to calculate the asymmetries for a certain produced hadron or hadron pair in the $Z^0$ decays. The difficulty in describing the weak interactions of quarks lies in the description of the quark fragmentation process. Thanks to the asymptotic freedom of QCD, the fragmentation process can be studied in the factorization theorem (see, e.g., Refs. \cite{Amati:1978wx,Ellis:1978ty,Collins:1989gx}). Factorization theorem tells that measurable quantities, e.g. cross section, can be separated by the calculable hard parts from the non-perturbative soft parts. If only the fragmentation process is taken into consideration, the non-perturbative soft parts are usually factorized as fragmentation functions (FFs) and/or di-hadron fragmentation functions (DiFFs). Quantities therefore can be expressed in terms of FFs and/or DiFFs in the annihilation and other fragmentation processes.

The discussion of the forward-backward asymmetry for a single certain hadron is available in Ref.  \cite{Yang:2022mmh}. In this paper, we consider the asymmetry for a hadron pair in the semi-inclusive electron positron annihilation process. The hadron pair is described by the DiFFs which were first introduced to describe the hadron pair production in a fragmenting jet at leading twist level in refs. \cite{Bianconi:1999cd,Bianconi:1999uc} and extended to the twist-3 level in Ref. \cite{Bacchetta:2003vn}. DiFFs are assumed to be universal and can be factorized in high energy reactions. By extracting from the two-jet events in the electron positron annihilation process \cite{Boer:2003ya,Bacchetta:2008wb,Courtoy:2012ry,Matevosyan:2018icf}, they can be used to study the nucleon structures in the framework of collinear factorization \cite{Bacchetta:2002ux,Bacchetta:2004it,Bacchetta:2011ip,Pisano:2015wnq}. Recent measurement of the invariant-mass dependence DiFF can be found in Ref. \cite{Belle:2017rwm}.
DiFFs are also considered to be strongly related to the jet handedness and can be used to investigate the quark and/or gluon polarizations \cite{Boer:2003ya,Pisano:2015wnq,Metz:2016swz}.
Here we note that the hadron pair have two origins. First of all, the pair could come from a single parton which fragments into two hadrons. Second, two hadrons can also arise when the parton first splits into two partons with each of them afterwards fragmenting into a single hadron. The key to distinguish the two possible origins is the size of the hadron pair invariant mass. If the invariant mass is much smaller than the hard scale, we can parametrize the fragmentation process with the nonperturbative DiFFs. Otherwise, we fall back to the convolution of two single-hadron FFs. For the first case, the DiFFs is similar to FFs and they have the same evolution equations \cite{Metz:2016swz,deFlorian:2003cg,Ceccopieri:2007ip}. For the second case, a splitting term should be added into the evolution equation. We do not consider this case in this paper.

Since factorization beyond leading order for twist-4 terms is unclear, we limit ourselves by leading order calculations in this paper. In other words, the calculations are carried out at leading order twist-4 by applying the collinear expansion formalism~\cite{Ellis:1982wd,Ellis:1982cd,Qiu:1990xxa,Qiu:1990xy}. This will greatly simplify the systematic calculation of higher twist contributions. The detailed calculations for the hadron pair production electron positron process by one of us can be found in \cite{Yang:2022knp}.  In this paper we use the results and introduce the definition of the forward-backward asymmetry for the hadron pair and present these asymmetry results in terms of DiFFs. With the available parameterizations of the DiFF from the Monte Carlo simulation \cite{Courtoy:2012ry}, we show numerical estimates for this asymmetry. We also present a numerical estimate of the twist-4 DiFFs in order to illustrate their contributions.
To be explicit, we organize the rest of the paper  as follows. In Sec.~\ref{sec:formalism}, we first introduce the general definition of the forward-backward asymmetry of the fermion pair in the annihilation process and show some conventions used in this paper. We also present the results of the hadron pair production process obtained in \cite{Yang:2022knp}. The explicit expressions of the forward-backward asymmetries in terms of DiFFs and numerical estimations are shown in Sec.~\ref{sec:results}. A brief summary is finally given in Sec.~\ref{sec:summary}.

\section{The differential cross section}\label{sec:formalism}

\subsection{The forward-backward asymmetry}

A simple exercise for the fermion pair production in the electron positron annihilation process is to calculate the muon pair production process. By considering the EW theory, the tree level differential cross section of this process can be written as
\begin{align}
  \frac{d\sigma}{d\cos\theta}=\frac{ \pi \alpha_{em}^2}{2Q^2}&\bigg\{\chi \left[c_1^ec_1^\mu \left(1+\cos^2\theta \right)+2c_3^ec_3^\mu \cos\theta \right]\nonumber\\
  +& \chi_{int}\left[c_V^ec_V^\mu \left( 1+\cos^2\theta \right)+2c_A^ec_A^\mu \cos\theta \right] \nonumber\\
  +& e_q^2\left(1+\cos^2\theta \right)\bigg\},  \label{f:cross}
\end{align}
where $\theta$ is the scattering angle in the lepton center-of-mass frame or the gauge boson rest frame, $\alpha_{em} = e^2/4\pi$ is the fine structure constant and $Q^2 = q^2=(l+l')^2$.
\begin{align}
   &\chi= \frac{Q^4}{\left[(Q^2-M_Z^2)^2 + \Gamma_Z^2 M_Z^2 \right] \sin^4 2\theta_W}, \\
   &\chi_{int} = -\frac{2e_q Q^2 (Q^2-M_Z^2)}{\left[(Q^2-M_Z^2)^2 + \Gamma_Z^2 M_Z^2 \right] \sin^2 2\theta_W},
\end{align}
where $M_Z$ and  $\Gamma_Z$ are respectively the mass and decay width  of $Z$-boson, $\theta_W$ is the weak mixing angle. $c_1^e = (c_V^e)^2 + (c_A^e)^2$ and $c_3^e = 2 c_V^e c_A^e$,
$c_V^e$ and $c_A^e$ are defined in the weak interaction current
$J_\mu (x)=\bar \psi(x)\Gamma_\mu\psi(x)$ where $\Gamma_\mu= \gamma_\mu (c_V^e - c_A^e \gamma^5)$.
Similar notations are also used for muon and quarks where we use a superscripts $\mu$ and $q$ to replace $e$.

The  forward-backward asymmetry in the angular distribution of positively and negatively charged fermions is defined as
\begin{align}
A_{FB} = \frac{\int_{0}^{1} d \sigma_\theta d \cos\theta -\int_{-1}^{0} d \sigma_\theta d \cos\theta  }{\int_{-1}^{1}   d \sigma_\theta d \cos\theta}, \label{f:fbdef}
\end{align}
where $d\sigma_\theta= d \sigma / d\cos\theta$ is the differential cross section given in Eq. (\ref{f:cross}). Using the definition in Eq. (\ref{f:fbdef}) and the differential cross section in Eq. (\ref{f:cross}), we obtain
\begin{align}
A^f_{FB} = \frac{3( \chi c_3^e c_3^f+\chi_{int} c_A^e c_A^f ) }{4(e_f^2+\chi c_1^e c_1^f + \chi_{int} c_V^e c_V^f  )}, \label{f:fbmnu}
\end{align}
where superscript $f$ denotes fermions (muon and quarks) and $e_f$ is the corresponding electric charge. In the following context, we extend the results to the hadron pair production process. We note that the definition of the forward-backward asymmetry given in Eq. (\ref{f:fbdef}) will be slightly modified for calculating that for hadron pairs. It will be shown in Sec. \ref{sec:results}.

%At the low-energy limit ($Q^2<<M_Z^2$), this asymmetry is given approximately by
%\begin{align}
% A^\mu_{FB} =- \frac{3G_F Q^2 c_A^e c_A^\mu }{4\sqrt{2}\pi \alpha_{em}},\label{f:fbmu}
%\end{align}
%where $G_F$ is the Fermi constant. Similar results can also be obtained for quarks as long as we replace the corresponding couplings for muon by that for quarks. For example, as the low-energy limit, we have
%\begin{align}
% A^q_{FB} = \frac{3G_F Q^2 c_A^e c_A^q }{4\sqrt{2}e_q\pi \alpha_{em}}, \label{f:fbq}
%\end{align}
%where $e_q$ is the electric charge of the quark with flavor $q$.  From Eqs. (\ref{f:fbmnu})-(\ref{f:fbq}), we can see that forward-backward asymmetries depend on weak couplings for certain fermions, they would give independent determinations of these couplings.

\begin{figure}
  \centering
 \includegraphics[width=0.4\linewidth]{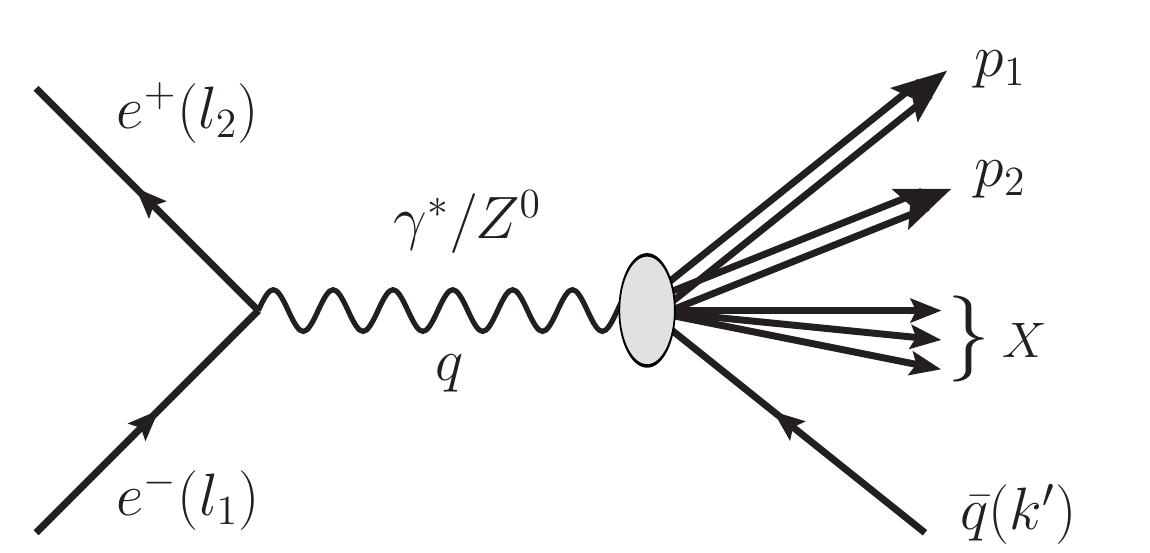}\\
  \caption{Illustrating diagram for the $e^+ + e^- \to h_1+h_2 + \bar q +X$ process.}\label{fig:dih}
\end{figure}

\subsection{The formalism}
To be explicit, we label the tree-level semi-inclusive electron positron annihilation process as
\begin{align}
e^+ + e^- \to q+\bar q \to h_1+h_2 + \bar q +X
\end{align}
where $\bar q$ denotes an antiquark that corresponds to a jet of hadrons and $h_1, h_2$ denote outgoing hadrons fragmented from $q$ in experiments, see Fig. \ref{fig:dih}. The differential cross section of this process is given by

\begin{align}
 d\sigma=&\frac{\alpha_{em}^2}{2\pi^2sQ^4}A_r L^r_{\mu\nu}(l_1,l_2)W_r^{\mu\nu}(p_1,p_2,k') \frac{d^3 p_1}{2E_1}\frac{d^3 p_2}{2E_2}\frac{d^3 k'}{2\pi 2E_k}, \label{f:cross1}
\end{align}
where $s=Q^2=q^2$ with $q=l_1+l_2$, $l_1, l_2$ are momenta of the leptons, $p_1,p_2$ are momenta of the outgoing hadrons.The symbol $r$ can be $\gamma\gamma$, $ZZ$ and $\gamma Z$, for electromagnetic (EM), weak and interference terms, respectively.
A summation over $r$ is understood, i.e., the total cross section is given by
\begin{align}
  %\frac{d\sigma}{dx dy d\psi d^2 k_\perp^\prime} = \frac{d\sigma^{ZZ} + d\sigma^{\gamma Z} + d\sigma^{\gamma\gamma}}{dx dy d\psi d^2 k_\perp^\prime}.
  d\sigma = d\sigma^{ZZ} + d\sigma^{\gamma Z} + d\sigma^{\gamma\gamma}. \label{f:threet}
\end{align}
$A_r$'s are defined as
\begin{align}
& A_{\gamma\gamma} = e_q^2, && A_{ZZ} = \chi, && A_{\gamma Z} =\chi_{int}.
\end{align}

The leptonic tensors for different cases are respectively given by
\begin{align}
%& L_{\gamma\gamma}^{\mu\nu}(l,\lambda_e, l^\prime) ={\rm Tr} \sum_{s^\prime}\left[ u_{s^\prime}(l^\prime)\bar u_{s^\prime}(l^\prime) \gamma^\nu u_{s}(l)\bar u_{s}(l) \gamma^\mu \right]
 &L^{\gamma\gamma}_{\mu\nu}(l_1, l_2)= l_{1\mu} l_{2\nu} +  l_{1\nu} l_{2\mu}  - (1_l\cdot l_2)g_{\mu\nu}, \\
 & L^{ZZ}_{\mu\nu}(l_1, l_2) =c_1^e \left[l_{1\mu} l_{2\nu} +  l_{1\nu} l_{2\mu}  - (1_l\cdot l_2)g_{\mu\nu}\right] +ic_3^e \varepsilon_{\mu\nu l_1l_2}, \\
 & L^{\gamma Z}_{\mu\nu}(l_1, l_2)=c_V^e \left[l_{1\mu} l_{2\nu} +  l_{1\nu} l_{2\mu}  - (1_l\cdot l_2)g_{\mu\nu}\right] \! + \!ic_V^e \varepsilon_{\mu\nu l_1l_2},
\end{align}
where coefficients $c_{1,3}^{e}$ have been introduced in the previous subsection.
%Similar notations are also used for quarks where the superscript $e$ is replaced by $q$.
The corresponding hadronic tensor are given by
\begin{align}
 W_{\gamma\gamma}^{\mu\nu}=&\sum_X\delta(q-p_1-p_2-k'-p_X) \langle 0 |J^\mu_{\gamma\gamma}(0) |p_1,p_2,k',X\rangle \langle p_1,p_2,k',X |J^\nu_{\gamma\gamma}(0) |0\rangle, \\
 W_{ZZ}^{\mu\nu}=&\sum_X\delta(q-p_1-p_2-k'-p_X) \langle 0 |J^\mu_{ZZ}(0)|p_1,p_2,k',X\rangle \langle p_1,p_2,k',X |J^\nu_{ZZ}(0) |0\rangle, \\
 W_{\gamma Z}^{\mu\nu}=&\sum_X\delta(q-p_1-p_2-k'-p_X) \langle 0 |J^\mu_{ZZ}(0)|p_1,p_2,k',X\rangle \langle p_1,p_2,k',X |J^\nu_{\gamma\gamma}(0) |0\rangle,
\end{align}
where $J^\mu_{\gamma\gamma}(0)=\bar \psi(0)\gamma^\mu\psi(0)$ and  $J^\mu_{ZZ}(0)=\bar \psi(0)\Gamma^\mu\psi(0)$.
Although, we have shown EM, weak and interference terms for both the leptonic and hadronic tensors, we only present  calculations of the weak interaction in the following context for simplicity. Other cases can be obtained in the similar way or by changing $c_1, c_3$ into $1, 0$ and $c_V, c_A$ for EM and interference cases, respectively.

%\begin{figure}
%  \centering
% \includegraphics[width=7cm]{frame}\\
%  \caption{Illustrating of the kinematics for the $e^+ + e^- \to h_1+h_2 + \bar q +X$ process.}\label{fig:frame}
%\end{figure}

In describing the hadron pair production in the electron positron annihilation process, following the previous conventions \cite{Bianconi:1999cd,Bianconi:1999uc}, we define $p_h=p_1+p_2$, $R=(p_1-p_2)/2$ and introduce the frame where momenta can be parameterized in the following forms:
\begin{align}
 & q=Q(1, 0, 0, 0), \\
 & l_1=\frac{Q}{2}(1,-\sin\theta, 0, -\cos\theta),\\
 &p_h=(E_h, 0, 0, p_z),\\
 & R=(E_r, | R_T|\cos\phi_r, | R_T|\sin\phi_r, R_z),\\
 &k'=(E_k, | k'_T|\cos\phi_k, | k'_T|\sin\phi_k, k'_z).
\end{align}
We also introduce the following standard variables used in this paper,
\begin{align}
& z=\frac{2p_h\cdot q}{Q^2}=\frac{2p_1\cdot q}{Q^2}+\frac{2p_2\cdot q}{Q^2}=z_1+z_2, \\
& \xi=\frac{z_1}{z}=1-\frac{z_2}{z}.
\end{align}
In terms of the variables above, the phase space factor can be rewritten as
\begin{align}
\frac{d^3 p_1}{2E_1}\frac{d^3 p_2}{2E_2}\frac{d^3 k'}{2E_k}=\frac{\pi}{8}zQ^2dzdy d\phi_r dM_h^2 d\xi \frac{dk'_z}{2E_k}d^2 k'_T.
\end{align}
Here we have used $d\Omega =2dy d\phi_L=4\pi dy$ where $y=p_h\cdot l_1/p_h\cdot q$, $\phi_L$ is the angle of lepton with respect to $p_h$. %$d^2R_T=\xi(1-\xi)d\phi_rdM_h^2$,  $M_h^2=p_h^2=(p_1+p_2)^2$.
The differential cross section therefore can be rewritten as
\begin{align}
 \frac{d\sigma}{dzdy d\phi_r M_h^2 d\xi d^2 k'_T}=\frac{\alpha_{em}^2 z}{16\pi Q^4}\chi L_{\mu\nu}(l_1,l_2) W^{\mu\nu}(p_h, k'_T). \label{f:crosssection}
\end{align}
Here the $k'_T$-dependent hadronic tensor is defined by integrating over $dk_z^\prime$,
\begin{align}
W^{\mu\nu}(p_h,  k'_T)=\int \frac{dk'_z}{2\pi 2E_k}W^{\mu\nu}(p_h, k'). \label{f:hadronict}
\end{align}
In the leptons center-of-mass frame $|k_T|=| k'_T|$, the hadronic tensor therefore can also be seen as a function of $k_T$.

%Hadronic tensor involving the nonperturbative information can not be calculated perturbatively.
%From the previous section, we see that the cross section is given by the contraction of the leptonic tensor and the hadronic tensor. To obtain the cross section, we need the explicit expression of the hadronic tensor in the parton model. We calculate the hadronic tensor in the following context.

\subsection{The cross section at twist-4}

Without repeating the tedious mathematical calculations which can be found in \cite{Yang:2022knp},  we here only write down the result of the differential cross section for simplicity. The complete expression for the weak interaction process at tree level twist-4 is given by
\begin{align}
 [d\sigma] &= \frac{\alpha_{em}^2\chi}{8\pi Q^2} \Bigg\{T_1(y) \left(D_1-\kappa^2_M\frac{D_{4q}}{z}\right) +T_2(y)k_{TM} R_{TM}\sin(\phi_r-\phi_k)\left(G^\perp_1-\kappa^2_M\frac{G_{4q}}{z}\right)  \nonumber\\
 &-2\kappa_M\left[ T_3(y)k_{TM}\cos\phi_k D^\perp + T_3(y)R_{TM}\cos\phi_r D^\sphericalangle \right]\nonumber\\
 &-2\kappa_M\left[ T_4(y)k_{TM}\sin\phi_k G^\perp + T_4(y)R_{TM}\sin\phi_r G^\sphericalangle \right] \nonumber\\
 &+4\kappa_M^2 \Big[2c_1^ec_1^q B(y)\frac{D_3}{z}+2c_1^ec_3^q B(y)k_{TM}R_{TM} \sin(\phi_r-\phi_k)\frac{G_3}{z}-T_1(y) \mathrm{Re}D_{-3dd} \Big] \nonumber\\
 &-4\kappa_M^2 c_1^ec_1^q B(y)\Big[k_{TM}^2\cos2\phi_k \mathrm{Re}D^\perp_{-3d}  + R_{TM}^2\cos2\phi_r \mathrm{Re}D^\sphericalangle_{-3d} + k_{TM}R_{TM}\cos(\phi_r+\phi_k) \mathrm{Re}D^\times_{-3d}\Big] \nonumber\\
 &-4\kappa_M^2  c_1^ec_3^q B(y)\Big[k_{TM}^2\sin2\phi_k \mathrm{Im}D^\perp_{-3d} + R_{TM}^2\sin2\phi_r \mathrm{Im}D^\sphericalangle_{-3d} + k_{TM}R_{TM}\sin(\phi_r+\phi_k) \mathrm{Im}D^\times_{-3d}\Big] \Bigg\}, \label{f:crossf}
\end{align}
where $[d\sigma]=d\sigma/dzdy d\phi_rdM_h^2 d\xi d^2 k'_T$. We also used $k_{TM}=| k_T|/M_h$, $R_{TM}=| R_T|/M_h$ and
\begin{align}
 &T_1(y)= 2c_1^ec_1^qA(y)-c_3^ec_3^qC(y), \\
 &T_2(y)= 2c_1^ec_3^qA(y)-c_3^ec_1^qC(y), \\
 &T_3(y)= c_1^ec_1^qC(y)D(y)+c_3^ec_3^qD(y), \\
 &T_4(y)= c_1^ec_3^qC(y)D(y)-c_3^ec_1^qD(y)
\end{align}
with $A(y)=\frac{1}{2}-y+y^2, B(y)=2y(1-y), C(y)=1-2y$ and $D(y)=\sqrt{y(1-y)}$ to simplify the expression. The twist suppression factor $\kappa_M$ is defined as $\kappa_M=M_h/Q$. Contributions from four-quark correlator which are labeled with subscript $4q$ are involved in Eq. (\ref{f:crossf}), e.g, $D_{4q}/z$. As for the EM and the interference cases, $T_i (i=1,2,3,4)$ are respectively reduced as
\begin{align}
 &T^{em}_1(y)= 2A(y), \\
 &T^{em}_2(y)= 0, \\
 &T^{em}_3(y)= C(y)D(y), \\
 &T^{em}_4(y)= 0,\\
%%%%%%%%%%%%
 &T^{int}_1(y)= 2c_V^ec_V^qA(y)-c_A^ec_A^qC(y), \\
 &T^{int}_2(y)= 2c_V^ec_A^qA(y)-c_A^ec_V^qC(y), \\
 &T^{int}_3(y)= c_V^ec_V^qC(y)D(y)+c_A^ec_A^qD(y), \\
 &T^{int}_4(y)= c_V^ec_A^qC(y)D(y)-c_A^ec_V^qD(y).
\end{align}
The kinematic factor $\chi$ should changes to $e_q^2$ and $\chi_{int}$ for EM and interference contributions, respectively. The total cross section is the sum of the EM, weak and interference terms, see Eq. (\ref{f:threet}).

\section{The results at twist-4} \label{sec:results}

\subsection{The forward-backward asymmetry}

The forward-backward asymmetry which is introduced to describe the angle distribution of the fermions from $Z^0$ decays has been introduced in Sec. \ref{sec:formalism}. Here we redefine the asymmetry at the hadonic level to illustrate the angle distribution of the produced hadron pair in the electron positron annihilation process. Comparing to Eq. (\ref{f:fbdef}), we define the forward-backward asymmetry for a hadron pair as
\begin{align}
 A_{FB}=\frac{\int_0^{1} [d\sigma] d\cos\theta-\int_{-1}^0 [d\sigma] d\cos\theta }{\int_{-1}^1 [d\sigma]_U d\cos\theta}, \label{f:FBtheta}
\end{align}
where $[d\sigma]$ is the differential cross section of the scattering angle $\theta (\cos\theta)$ while $[d\sigma]_U$ denotes the azimuthally independent one.
%This definition is different from that in Eq.  (\ref{f:fbdef}) where the interference and weak terms are also included in the denominator.  However, as the low-energy limit, the electromagnetic term dominates, see Eqs. (\ref{f:fbmu}) and (\ref{f:fbq}).
Instead of $\cos\theta$, the differential cross section is given in terms of $y$ in Eq. (\ref{f:crossf}). It is then convenient to rewrite the forward-backward asymmetry $A_{FB}$ in the following form,
\begin{align}
 A_{FB}=\frac{\int_{1/2}^{1} (d\sigma) dy-\int_{0}^{1/2} (d\sigma) dy }{\int_{0}^{1} (d\sigma)_U dy}, \label{f:FBy}
\end{align}
where  $(d\sigma)=d\sigma/dzdy d\phi_r dM_h^2 d\xi d^2 k'_T$  and $ (d\sigma)_U$ only includes azimuthally independent terms.  Remember the differential cross section is a sum of the EM, weak and interference terms, see Eq. (\ref{f:threet}). Then we obtain the denominator:
\begin{align}
 \int^1_0  (d\sigma)_U dy= \frac{\alpha_{em}^2}{12 Q^2}\left(e_q^2+\chi c_1^e c_1^q + \chi_{int} c_V^e c_V^q \right) \tilde D_1, \label{f:crossunp}
\end{align}
where $\tilde D_1 =D_1- \kappa^2_M D_{4q}/z $.

Using the asymmetry definition in Eq. (\ref{f:FBy}) and the corresponding differential cross section, we obtain 6 kinds of asymmetries, 2 of them are leading twist effects while 4 of them are twist-3 effects:
\begin{align}
  A_{FB}&=\frac{3\left(\chi c_3^e c_3^q +\chi_{int} c_A^e c_A^q  \right)\tilde D_1}{4\left(e_q^2+\chi c_1^e c_1^q +\chi_{int} c_V^e c_V^q  \right)\tilde D_1},  \label{f:fbaun} \\
  A_{FB}^{kR,y}&=\frac{3\left(\chi c_3^e c_1^q +\chi_{int} c_A^e c_V^q  \right)\tilde G^\perp_1}{4\left(e_q^2+\chi c_1^e c_1^q +\chi_{int} c_V^e c_V^q  \right)\tilde D_1},  \label{f:fbakrx} \\
  A_{FB}^{k,x}&=- k_{T M}\frac{\kappa_M\left(\chi c_1^e c_1^q +\chi_{int} c_V^e c_V^q  \right) D^\perp}{2\left(e_q^2+\chi c_1^e c_1^q +\chi_{int} c_V^e c_V^q  \right)\tilde D_1},  \label{f:fbakx} \\
  A_{FB}^{k,y}&=- k_{T M}\frac{\kappa_M\left(\chi c_1^e c_3^q +\chi_{int} c_V^e c_A^q  \right)G^\perp}{2\left(e_q^2+\chi c_1^e c_1^q +\chi_{int} c_V^e c_V^q  \right)\tilde D_1},  \label{f:fbaky} \\
  A_{FB}^{r,x}&=- R_{T M}\frac{\kappa_M\left(\chi c_1^e c_1^q +\chi_{int} c_V^e c_V^q  \right) D^\sphericalangle}{2\left(e_q^2+\chi c_1^e c_1^q +\chi_{int} c_V^e c_V^q  \right)\tilde D_1},  \label{f:fbarx} \\
  A_{FB}^{r,y}&=- R_{T M}\frac{\kappa_M\left(\chi c_1^e c_3^q +\chi_{int} c_V^e c_A^q  \right)G^\sphericalangle}{2\left(e_q^2+\chi c_1^e c_1^q +\chi_{int} c_V^e c_V^q  \right)\tilde D_1}, \label{f:fbary}
\end{align}
where $\tilde G^\perp_1 =G^\perp_1- \kappa^2_M G_{4q}/z $.
Twist-4 asymmetries are suppressed at high energies, we do not show them here. We note that a summation of flavor $q$  is explicit in the numerator and in the denominator, respectively. This applies also to all the results presented in the following context.
From the point of view of experimental measurement, it is convenient to consider contributions from the collinear DiFFs.
If one integrates over $k_T$, only asymmetries shown in Eqs. (\ref{f:fbaun}), (\ref{f:fbarx}) and (\ref{f:fbary}) are left.

Considering the low energy limit, forward-backward asymmetries shown above can be rewritten as
\begin{align}
  A_{FB}&=\frac{3G_F Q^2  c_A^e c_A^q \tilde D_1}{4\sqrt 2 e_q \pi \alpha_{em}\tilde D_1},  \label{f:fbaunlow} \\
  A_{FB}^{kR,y}&=\frac{3G_F Q^2  c_A^e c_V^q \tilde G^\perp_1}{4\sqrt 2 e_q \pi \alpha_{em}\tilde D_1},  \label{f:fbakrxlow} \\
  A_{FB}^{k,x}&=-k_{T M}\kappa_M\frac{G_F Q^2  c_V^e c_V^q  D^\perp}{2\sqrt 2 e_q \pi \alpha_{em}\tilde D_1},  \label{f:fbakxlow} \\
  A_{FB}^{k,y}&=-k_{T M}\kappa_M\frac{G_F Q^2  c_V^e c_A^q  G^\perp}{2\sqrt 2 e_q \pi \alpha_{em}\tilde D_1},    \label{f:fbakylow} \\
  A_{FB}^{r,x}&=-R_{T M}\kappa_M\frac{G_F Q^2  c_V^e c_V^q  D^\sphericalangle}{2\sqrt 2 e_q \pi \alpha_{em}\tilde D_1}, \label{f:fbarxlow} \\
  A_{FB}^{r,y}&=-R_{T M}\kappa_M\frac{G_F Q^2  c_V^e c_A^q  G^\sphericalangle}{2\sqrt 2 e_q \pi \alpha_{em}\tilde D_1}, \label{f:fbarylow}
\end{align}
where $G_F=g^2/4\sqrt 2 M_W^2 =e^2/\sqrt 2M_Z^2 \sin^22\theta_W$, is the Fermi constant.

\begin{figure}
  \centering
  \includegraphics[width=0.5\linewidth]{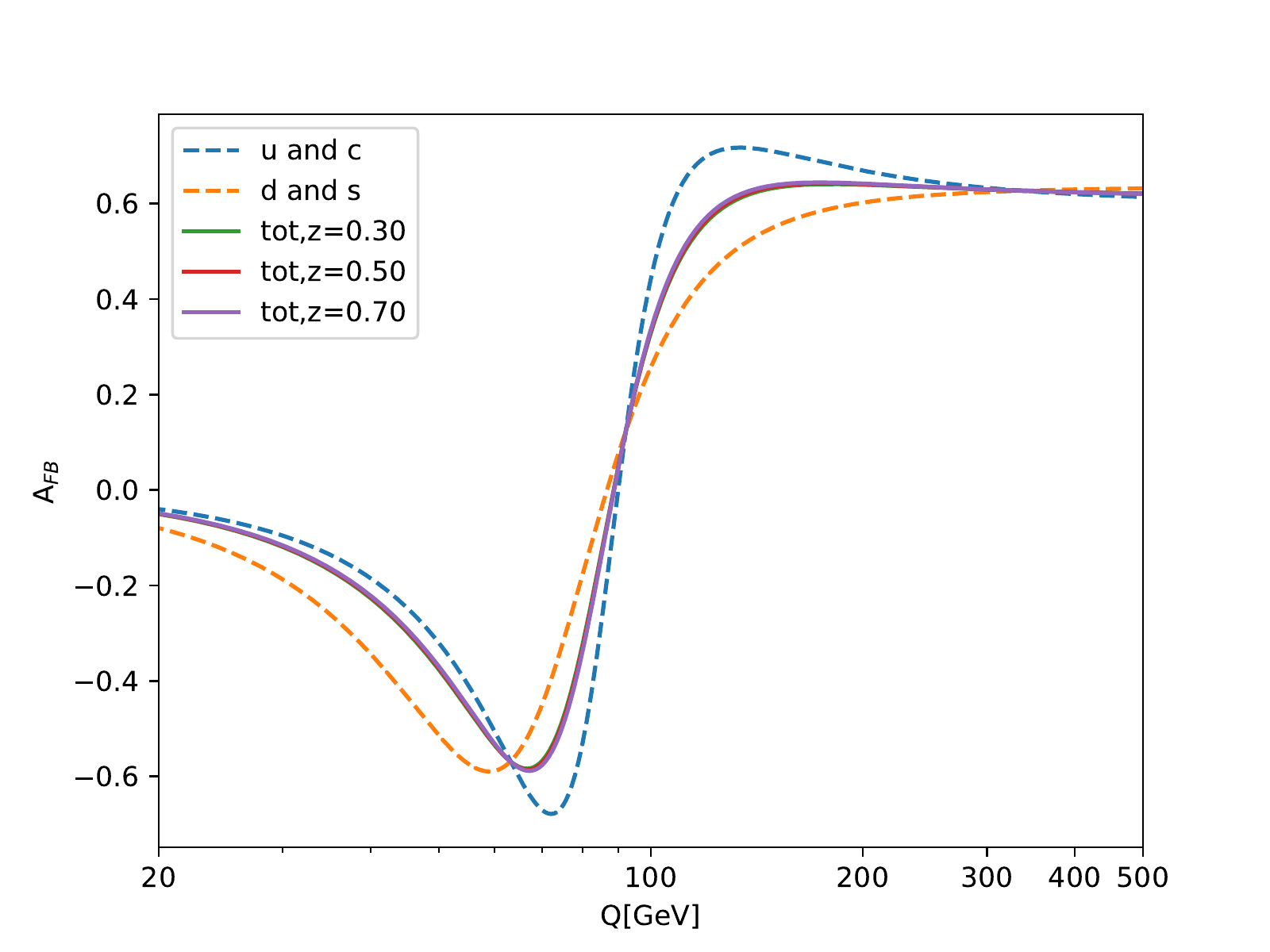}
  \caption{The forward-backward asymmetry for $A_{FB}$. Dashed lines show asymmetries for quarks. Solid lines show asymmetries for produced hadron pairs with different momentum fraction $z$. Here $M_h=0.5 GeV$.} \label{AFB}
\end{figure}

To have an intuitive impression of the hadron forward-backward asymmetry shown above, we present the numerical values of $ A_{FB}$ in Fig. \ref{AFB}. We do not show other asymmetries due to lack of proper parameterizations. Parameterizations of $D_1(z, M_h, Q^2)$ are taken from ref. \cite{Courtoy:2012ry} in which only $u, d, s$ and $ c$ quarks are considered. The QCD evolution of the DiFF starts from $Q^2 = 1 GeV$ and is limited at leading
order \cite{Metz:2016swz,deFlorian:2003cg,Ceccopieri:2007ip}. Invariant mass $M_h$ is chosen as $0.5 GeV$. As a contrast, the asymmetries for quarks are shown by dashed lines while asymmetries for the produced hadron pair are shown by solid lines. From Fig. \ref{AFB}, we see the forward-backward asymmetry of the hadron pair is insensitive to the momentum fraction $z$ but can be distinguished from that of the quarks. %It is also interesting to see the asymmetry of the $M_h$-dependence. We show the asymmetries in Fig. \ref{AFBMh}. The invariant mass is chosen as $M_h=0.5 GeV$, $0.8 GeV$, $1.0 GeV$ and $z=0.5$. We can see that the forward-backward asymmetry for the hadron pair is also insensitive to $M_h$.

%\begin{figure}
%  \centering
%  \includegraphics[width=0.5\linewidth]{AFBMh}
%  \caption{The forward-backward asymmetry for $A_{FB}$ with different $M_h$. Asymmetries for quarks are neglected. Here $z=0.5$.}\label{AFBMh}
%\end{figure}

\subsection{Estimation of the twist-4 DiFFs}

In this paper, our calculations are shown at twist-4 level. In order to estimate the contributions from the twist-4 DiFFs, we introduce the ratio of the twist-4 DiFFs to the leading twist DiFF. For simplicity, we consider
\begin{align}
 [d\sigma]&= \frac{\alpha_{em}^2\chi}{8 Q^2} \Bigg\{T_1(y) \left(D_1-\kappa^2_M\frac{D_{4q}}{z}\right) +4\kappa_M^2 \Big[2c_1^ec_1^q B(y)\frac{D_3}{z}-T_1(y) \mathrm{Re}D_{-3dd} \Big]
 \Bigg\},
\end{align}
where $k_T$- and $R_T$-dependent terms are neglected. In terms of twist-4 contribution factor or the ratio $\alpha$, we have
\begin{align}
 [d\sigma]&= \frac{\alpha_{em}^2\chi}{8 Q^2} \left[T_1(y) D_1 \left(1+\kappa^2_M \alpha \right)-T_1(y)\kappa^2_M\frac{D_{4q}}{z}  \right],
\end{align}
where
\begin{align}
  \alpha =4\frac{2c_1^e c_1^q B(y)D_3 - z T_1(y)\mathrm{Re} D_{-3dd}}{z T_1(y)D_1}.
\end{align}
We next need to find the relations between $D_3, \mathrm{Re} D_{-3dd}$ and $D_1$. Since $D_1$ and $D_3$ can be obtained by calculating the traces from the decomposition of the correlation function, using $\slashed k =k^+\slashed{\bar n} + k^-\slashed n + \slashed k_T$, we finally have
\begin{align}
 & D_3=-\frac{z^2 k_T^2}{2M_h^2}D_1. \label{f:D1D3}
% & 2\mathrm{Re} D_{3dd}=\frac{z^2k_T^2}{2M_h^2}\frac{\partial }{\partial z}D_1. \label{f:D1D3dd}
\end{align}
By using the rough approximation $g=0$, i.e., neglecting the contribution from gluons, we calculate the quark-gluon-gluon-quark correlator and obtain \cite{Yang:2022knp}
\begin{align}
  \big(\hat\Xi^{(2)}_{\rho\sigma}+\gamma^0\hat\Xi^{(2)\dag}_{\sigma\rho}\gamma^0\big)|_{g=0}=k_{T\rho}k_{T\sigma} z^2 \frac{\partial }{\partial z} \hat \Xi^{(0)}|_{g=0}. \label{f:decomg0}
\end{align}
Inserting the decomposition of the quark-j-gluon-quark correlator into Eq. (\ref{f:decomg0}) and neglecting the T-odd functions (e.g., $G_{3dd}$), we obtain
\begin{align}
% & D_3=-\frac{z^2 k_T^2}{2M_h^2}D_1. \label{f:D1D3}
 & 2\mathrm{Re} D_{3dd}=\frac{z^2k_T^2}{2M_h^2}\frac{\partial }{\partial z}D_1. \label{f:D1D3dd}
\end{align}
Therefore the $\alpha$ can be rewritten as
\begin{align}
  \alpha = zk_{TM}^2 \left[\frac{\partial \ln T_1(y)D_1}{\partial \ln z} + \frac{4c_1^e c_1^q B(y) D_1}{T_1(y)D_1}\right]. \label{f:alpha}
\end{align}

\begin{figure}
  \centering
  \includegraphics[width=0.5\linewidth]{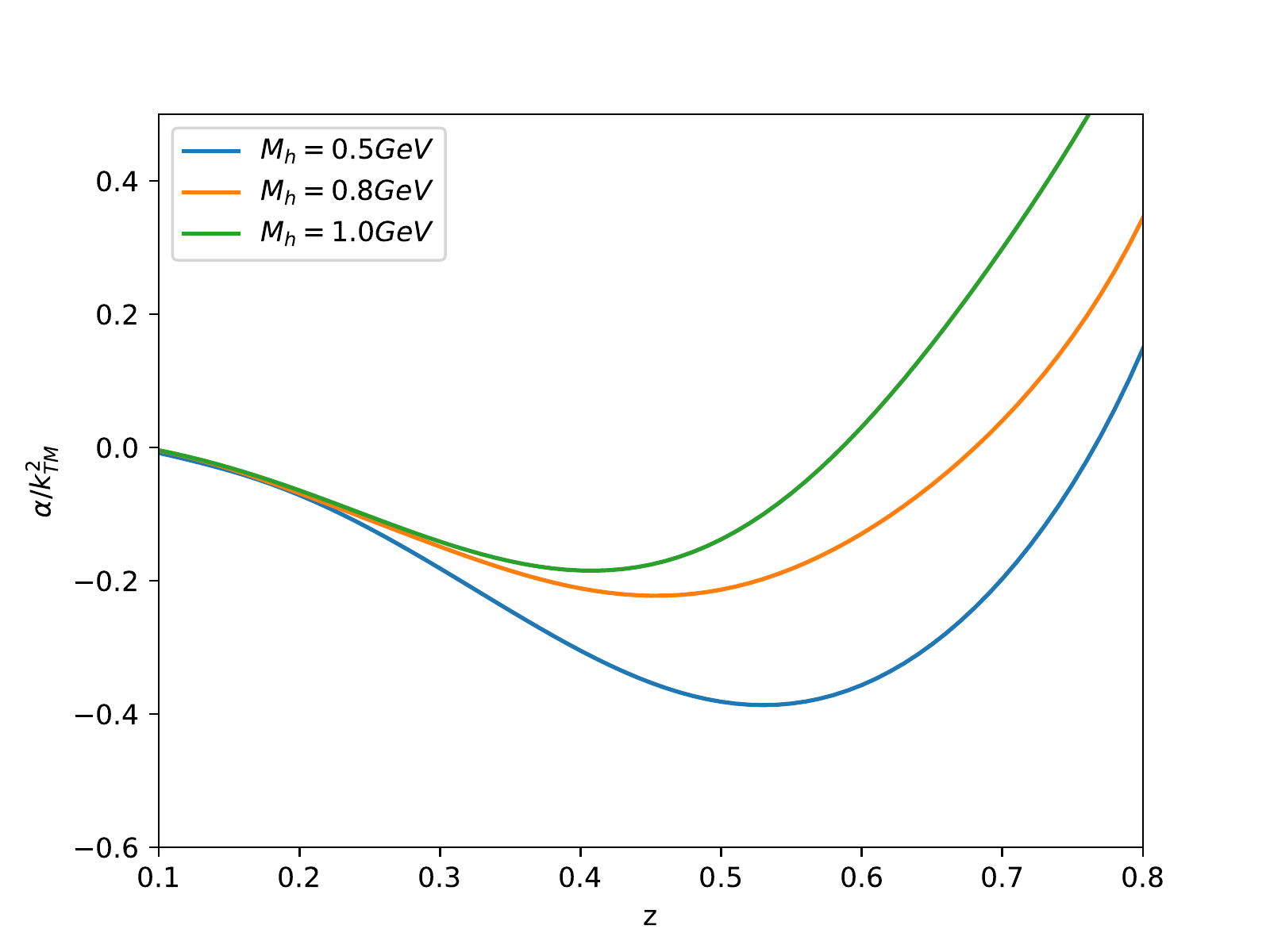}
  \caption{A rough estimation of the twist-4 contribution factor $\alpha/k_{TM}^2$ as a function of $z$ at $Q=M_Z$ with different $M_h$.}\label{alpha}
\end{figure}

With the parametrization of $D_1$ used in the previous subsection, we show a rough estimation of the twist-4 contribution factor $\alpha$ in Fig. \ref{alpha}.  We can see that the ratio $\alpha$ is sensitive to the invariant mass of the hadron pair for $z>0.3$ and the twist-4 DiFFs have approximately the same order of magnitude as the leading twist DiFF $D_1$. Higher twist suppressions come from the factor $\kappa_M$ rather than the higher twist parton functions. However, for $z<0.3$, the twist-4 DiFFs are one order of magnitude smaller than the leading twist DiFF $D_1$.

\subsection{The azimuthal asymmetries}

From Eq. (\ref{f:crossf}), we can see there are sets of azimuthal modulations which can be measured in experiments and used to extract the corresponding DiFFs. We show the results for completeness in the following.

To clarify our calculations we first present the definition of the azimuthal asymmetries, e.g.
\begin{align}
 \langle \sin\phi_k \rangle =\frac{\int [d\sigma]\sin\phi_k d\phi_k}{\int [d\sigma] d\phi_k}.
\end{align}
Other asymmetries can be defined in the similar way, we do not show them here. According to the definition, we write down all the azimuthal asymmetries. The leading twist asymmetry is given by
\begin{align}
 & \langle \sin(\phi_r-\phi_k) \rangle_2 =k_{TM}R_{TM}\frac{\chi T_2(y)G^\perp_1}{2\chi T_1(y)D_1}. \label{f:leadingasy}
\end{align}
Here subscript $2$ denotes the leading twist. The twist-4 correction of the leading twist asymmetry in Eq. (\ref{f:leadingasy}) in the numerator is $\kappa_M^2\left(-T_2(y)G_{4q}+8c_1^ec_3^qB(y)G_3\right)/z$.

There are four twist-3 azimuthal asymmetries which are given by
\begin{align}
& \langle \cos\phi_k \rangle_3 =-\kappa_{M}k_{TM}\frac{\chi T_3(y)D^\perp}{z\chi T_1(y)D_1}, \label{f:t3cosk} \\
& \langle \cos\phi_r \rangle_3 =-\kappa_{M}R_{TM}\frac{\chi T_3(y)D^\sphericalangle}{z\chi T_1(y)D_1}, \label{f:t3cosr} \\
& \langle \sin\phi_k \rangle_3 =-\kappa_{M}k_{TM}\frac{\chi T_4(y)G^\perp}{z\chi T_1(y)D_1}, \label{f:t3sink} \\
& \langle \sin\phi_r \rangle_3 =-\kappa_{M}R_{TM}\frac{\chi T_4(y)G^\sphericalangle}{z\chi T_1(y)D_1}, \label{f:t3sinr}
\end{align}
where subscript $3$ denotes the twist-3. There are six twist-4 azimuthal asymmetries which are given by
\begin{align}
& \langle \cos2\phi_k \rangle_4 =-\kappa^2_{M}k^2_{TM}\frac{2\chi c_1^ec_1^qB(y)\mathrm{Re}D_{-3d}^\perp}{z\chi T_1(y)D_1}, \label{f:t4cos2k} \\
& \langle \cos2\phi_r \rangle_4 =-\kappa^2_{M}R^2_{TM}\frac{2\chi c_1^ec_1^qB(y)\mathrm{Re}D_{-3d}^\sphericalangle}{z\chi T_1(y)D_1}, \label{f:t4cos2r} \\
& \langle \sin2\phi_k \rangle_4 =-\kappa^2_{M}k^2_{TM}\frac{2\chi c_1^ec_3^qB(y)\mathrm{Im}D_{-3d}^\perp}{z\chi T_1(y)D_1}, \label{f:t4sin2k} \\
& \langle \sin2\phi_r \rangle_4 =-\kappa^2_{M}R^2_{TM}\frac{2\chi c_1^ec_3^qB(y)\mathrm{Im}D_{-3d}^\sphericalangle}{z\chi T_1(y)D_1}, \label{f:t4sin2r} \\
& \langle \cos(\phi_r+\phi_k) \rangle_4 =\kappa^2_{M}k_{TM}R_{TM}\frac{2\chi c_1^ec_1^qB(y)\mathrm{Re}D_{-3d}^\times}{z\chi T_1(y)D_1}, \label{f:t4cos2kr} \\
& \langle \sin(\phi_r+\phi_k) \rangle_4 =\kappa^2_{M}k_{TM}R_{TM}\frac{2\chi c_1^ec_3^qB(y)\mathrm{Im}D_{-3d}^\times}{z\chi T_1(y)D_1}, \label{f:t4sin2rr}
\end{align}
where subscript $4$ denotes the twist-4.

\section{Summary}\label{sec:summary}

The forward-backward asymmetry is an important measurable quantity which can be used to  determine the neutral-current couplings. In this paper, we extend the definition of the asymmetry from the partonic level to the hadronic level by calculating this asymmetry of the hadron pair in the semi-inclusive electron position annihilation process. The production of the hadron pair is described by the DiFFs. Semi-inclusive implies that a back-to-back jet is also measured in addition to the hadron pair. We first calculate the cross section according to the collinear expansion method. It provides explicit expressions of the hadronic tensor at twist-4 level and the cross section can be easily obtained. We calculate the leading twist and twist-3 forward-backward asymmetries and present the numerical estimates at leading twist with available parametrizations. We find that the forward-backward asymmetry of the hadron pair is insensitive to the momentum fraction $z$ but can be distinguished from that of the quarks.
We also show a rough estimation of the twist-4 DiFFs. We find that the twist-4 DiFFs are one order of magnitude smaller than the leading twist DiFF $D_1$ in the $z<0.3$ region. However, for $z>0.3$, the twist-4 DiFFs have approximately the same order of magnitude as the leading twist DiFF $D_1$.
Azimuthal asymmetries are also shown for completeness.

%In this paper, we calculate the forward-backward asymmetries of the hadron pair in the semi-inclusive electron positron annihilation process at twist-4 level. Semi-inclusive implies the back-to-back jet is also measured in addition to the hadron pair. This process (jet production) is better than the double hadron (pair) production process because it does not introduce the extra uncertainties if the jet is seen as a(n) (anti)quark. It is then an ideal place to study the chiral even quantities (e.g. DiFFs). However, the shortcoming of this process is that it is impossible to study the chiral odd quantities since there is no helicity flip. Both the EM and weak interactions are considered.

\section*{Acknowledgments}
This work was supported by the Natural Science Foundation of Shandong Province (No. ZR2021QA015).

 \end{document}